\documentclass[epj]{svjour}
\usepackage{graphics}
\usepackage[utf8]{inputenc}
\usepackage[T1]{fontenc}
\usepackage{amsmath}
\usepackage{mathtools}
\usepackage{amsfonts}
\usepackage{amssymb}
\usepackage[utf8]{inputenc}
\usepackage{graphicx}
\usepackage{float}
\usepackage{latexsym}
\usepackage{enumitem}
\usepackage{subfigure}
\usepackage{epsfig}
\usepackage{rotating}
\usepackage{ragged2e}
\usepackage{mathtools}
\usepackage{tikz}
\usepackage{hyperref}
\usepackage{multirow}
\usepackage{cite}
\usepackage{widetext}
\newcommand{\orcid}[1]{\href{https://orcid.org/#1}{\includegraphics[width=8pt]{orcid.svg}}}

\begin{document}
\title{Electric Penrose process and collisions of particles near five-dimensional weakly charged Schwarzschild black hole}
\author{Tursunali Xamidov\inst{1,}\thanks{xamidovtursunali@gmail.com}, Mirzabek Alloqulov \inst{2,3,}\thanks{malloqulov@gmail.com}, Sanjar Shaymatov  \inst{1,4,5,6,}\thanks{sanjar@astrin.uz}}

\institute{Institute of Fundamental and Applied Research, National Research University TIIAME, Kori Niyoziy 39, Tashkent 100000, Uzbekistan \and University of Tashkent for Applied Sciences, Str. Gavhar 1, Tashkent 100149, Uzbekistan \and Shahrisabz State Pedagogical Institute, Shahrisabz Str. 10, Shahrisabz 181301, Uzbekistan \and Tashkent State Technical University, Tashkent 100095, Uzbekistan \and Samarkand State University, University Avenue 15, 140104 Samarkand, Uzbekistan %Institute for Theoretical Physics and Cosmology, Zhejiang University of Technology, Hangzhou 310023, China 
\and Western Caspian University, Baku AZ1001, Azerbaijan}
\date{Received: date / Revised version: date}
\abstract{
The particle dynamics and the electric Penrose process for the five-dimensional weakly charged Schwarzschild black hole are studied. Firstly, the horizon structure and the effective potential for the test particle are explored. The radial profile of the effective potential is plotted for different values of the BH charge. Then, we studied energy efficiency using the conservation laws for energy, angular momentum, and charge of particles. The appropriate plots were obtained and compared with the results for the four-dimensional Schwarzschild BH. Moreover, we obtained the constraints on mass and charge of black hole to accelerate protons of different energies ($1$-$10^{9}$ GeV). Finally, collisions of electrically charged particles near the horizon of the five-dimensional Schwarzschild BH are studied. We demonstrated the radial dependence of the center of mass energy for different values of the angular momentum and charge of the particles together with the BH charge. 
}
\maketitle

\section{Introduction}
\label{introduction}

In general relativity (GR) black holes are believed to play an important role in highly energetic astrophysical events such as $\gamma$-ray bursts~\cite{Meszaros06}, active galactic nuclei (AGN)~\cite{Peterson:97book}, and ultraluminous X-ray binaries~\cite{King01ApJ}. Therefore, black holes can be considered enormous energy reservoirs in the Universe. Therefore, the energetics are of much interest and can be considered a potential explanation for highly energetic astrophysical phenomena. As a result, some models for the energetics of black holes occurred. The first of them was formulated by Penrose~\cite{Penrose:1969pc} which is about extracting energy from rotating black holes. According to the Penrose process, the rotational energy of the black holes can be extracted by particle fission ($0 \rightarrow 1+2$) which occurs in the ergosphere that appears in the region between the horizon and the static limit bounded from the outer surface. Over the years, alternative mechanisms of energy extraction processes have been formulated, such as the collisional Penrose process~\cite{1975ApJ...196L.107P}, the Blandford-Znajek process~\cite{Blandford1977}, the magnetic Penrose process~\cite{Bhat85}, and the magnetic reconnection process~\cite{PhysRevD.103.023014,PhysRevD.110.044005} etc.

As mentioned, a falling particle splits into two pieces, that is, the one falls into black hole, while the second piece goes back to infinity, carrying out more energy than the initial energy. This how the escaping particle can take energy out from the black hole, resulting in the black hole's rotation slowing down. This process has been considered in different situations (see, for example, \cite{Abdujabbarov11,Toshmatov:2014qja,Okabayashi20,Xamidov24EPJC}). The Penrose process has also been applied in the context of Buchdahl stars \cite{Shaymatov24MPP2} and higher dimensional rotating black holes \cite{Prabhu10,Nozawa05,Shaymatov24MPP1}. The extraction of more energy from black hole strongly depends on the case for which the incident particle must be relativistic \cite{Bardeen72,Wald74ApJ}. For that, the PP was reformulated as the magnetic Penrose process, which allows an escaping particle to surpass the constraint velocity for being relativistic. This mechanism significantly enhances the efficiency of energy extraction from rotating black holes \cite{Bhat85,Parthasarathy86}. This mechanism addresses purely magnetic field effects on the energy extraction process from black holes. On these lines, an extensive analysis has since been implemented in various contexts    \cite{Wagh89,Alic12ApJ,Moesta12ApJ,Dadhich18mnras,Tursunov:2019oiq,Tursunov20ApJ,Shaymatov22b}). It is important to note that the energy extraction process through these mechanism can require the presence of the ergoregion and can only take place inside this region of rotating black holes. However, the Penrose process can also be extended to non-rotating black holes, i.e., the electric Penrose process using electric field to serve as a high-energy emission astrophysical phenomena (see, for example, \cite{Denardo73PLB,Tursunov21EPP,Stuchlik:2021,Alloqulov23EPP,Baez2024EPP,Alloqulov24EPP,Kurbonov2023}). 

One also needs to explore a proposed mechanism as a potential explanation for the aforementioned highly energetic astrophysical events. In this regard, high energy collisions of particles were theoretically formulated by Banados, Silk and West (BSW) \cite{Banados2009PRL}, considering a Kerr black hole as a particle accelerator. In this mechanism, it is envisioned that the center-of-mass energy can infinitely increase provided that the collision of two particles can occur near the horizon of an extremal Kerr black hole with maximal rotation. This has sparked increased research activity, focused on analyzing the the energy extraction from black holes and generalizing the BSW mechanism to various gravity models \cite{Banados2009PRL,Jacobson2010PRL,Lake2010PRL,Kimura2010PRD,Atamurotov13a,Atamurotov2022,Shaymatov13,Grib2011AP,Grib2011GC,Patil2010PRD,Patil2011PRD,Patil2011PRDa,Patil2011CQG,Harada2011PRD,Shaymatov18a,Harada2011PRDa,Shaymatov21pdu,Shaymatov21c}. 
 
In recent years, black hole solutions in spacetimes with more than four dimensions particularly in five dimensions have been a focal point of research, driven by advancements in braneworld concepts, string theory, and gauge/gravity duality. Numerous intriguing and surprising results have emerged in~\cite{Horowitz2011}. Note that the uniqueness theorems fail to apply to high dimensions because there are more degrees of freedom. From this perspective, the vacuum solution of GR for a five-dimensional, static, and spherically symmetric BH is the Schwarzschild-Tangherlini BHs solution~\cite{Tangherlini1963}. It is also important to note that there exist various supergravity solutions describing a higher dimensional charged black hole solution with a slow rotation limit~\cite{Aliev06,Aliev07a} and charged rotating black hole solutions in the context of supergravity and string theory \cite{Cvetic96,Chong05a}. An extensive analysis has been conducted in Refs.~\cite{Aman06,Sharif16,Xie22PLB} addressing the properties of a static spherically symmetric higher-dimensional Reissner-Nordström (RN) black hole, including its thermodynamic curvatures, scalarization and accretion properties around a higher-dimensional RN black hole. 

Taken together, it is also important to examine the energetic properties of a static spherically symmetric higher-dimensional black hole to gain a deeper explanation of high-energy astrophysical phenomena. This paper aims to investigate the energetics of a five-dimensional weakly charged Schwarzschild BH. Particularly, the electric Penrose process for five-dimensional Schwarzschild BH and collisions of electrically charged particles near the event horizon of the five-dimensional Schwarzschild BHs are investigated. It is worth noting that black holes with electric charge are an intriguing class of solutions in general relativity and higher-dimensional gravity theories. The most well-known charged black hole solution is the Reissner-Nordström (RN) metric, which describes a spherically symmetric black hole with an electrostatic charge. While such solutions are mathematically well-defined, their astrophysical relevance remains uncertain. Observational evidence suggests that realistic black holes are at most weakly charged. Therefore, most black holes in astrophysical environments are expected to be effectively uncharged or only weakly charged. Moreover, we compared our findings with known results for 4D Schwarzschild BHs~\cite{Tursunov21EPP}.

In Sec.~\ref{Sec:Sw5D} we describe briefly the five-dimensional weakly charged black hole and dynamics of the charged particles. In
Sec.~\ref{Sec:EP} describes the electric Penrose process for the five-dimensional weakly charged Schwarzschild black hole. In Sec.~\ref{Sec:energy} we investigate the collisions of electrically charged particles near the event horizon of the 5D Schwarzschild black holes. We summarize our concluding
remarks of the obtained results in Sec.~\ref{Sec:Conclusion}.
Throughout the paper, we use a system of units in which $G=c=1$, unless the constants are directly stated.

\section{Five dimensional weakly charged Schwarzschild black hole }\label{Sec:Sw5D}

The metric of a high-dimensional Schwarzschild black hole is expressed as follows \cite{MyersPerry1986, Aman06}
\begin{equation}\label{eq:5dmetric}
    d s^2 = -f(r) dt^2 + f(r)^{-1} dr^2  + r^2 d\theta^2 +
      r^2d\Omega^2_{D-2} \ ,
\end{equation}
where the metric function $f(r)$ and the line element of $D-2$ dimensional unit sphere $d\Omega^2_{D-2}$ are given, respectively, by
\begin{equation}\label{eq:5DmetricFunction}
    f(r) = 1- \frac{M}{r^{D-3}} \ ,
\end{equation}
and
\begin{equation}
    d\Omega^2_{D-2} = d\theta_1^2 + \sin^2\theta_1 \, d\theta_2^2 + \cdots + \left( \prod_{j=1}^{D-3} \sin^2\theta_j \right) d\theta_{D-2}^2 .
\end{equation}
For $D = 5$, the above equation takes the following form
\begin{equation}
    d\Omega^2_{3} = d\theta^2 + \sin^2\theta \left(d\phi^2 + \sin^2\phi \, d\chi^2\right) ,
\end{equation}
where the coordinates vary within the ranges $0 \leq \theta \leq \pi $, $ 0 \leq \phi \leq \pi $, and $ 0 \leq \chi \leq 2\pi $ \cite{MyersPerry1986, Aman06}.
In this research, we study a non-rotating, weakly charged five dimensional Schwarzchild black hole. We assume that the charge  $Q$ of the black hole is small enough to not affect the metric and located at the center of the coordinate system. In this case, the only non-zero component of the electromagnetic four-potential is
\begin{eqnarray} \label{el:pot} 
A_t = -\frac{Q}{r^{D-3}}\ . 
\end{eqnarray} 
The horizon $r_h$ of $D$-dimensional Schwarzschild black hole can be defined as follows
\begin{equation}
    r_h = \left( \frac{GM}{c^2}\right)^\frac{1}{D-3} .
    \label{eq:horizon}
\end{equation}
 For five dimensional black hole, the equation takes the following form
\begin{equation}
    r_h = \sqrt{\frac{GM}{c^2}} .
    \label{eq:5dhorizon}
\end{equation}
Let us suppose a charged particle is falling into the five dimensional Schwarzchild black hole. According to the symmetries in the system, we can write the following integrals of motion for this system
\begin{equation} \label{int: motion-energy}
   p_t = -E = mu_t + q A_t ,
\end{equation}
\begin{equation} \label{int: motion-ang-mom}
    p_\chi = L =m u_\chi ,
\end{equation}
where $p_t$ and $p_\chi$ are $t$ and $\chi$ components of the canonical five-momentum.

The radial motion of the particle can be described by its effective potential. Since this type of problems has been extensively studied by others, we will focus solely on the expression for the effective potential $V_{eff}(r)$ in this system
\begin{equation} \label{five:veff}
    V_{eff}(r)=\frac{\beta}{r^{2}}+\sqrt{\left(1-\frac{M}{r^2 }\right)\left(1+\frac{\mathcal{L}^2}{r^2}\right)}\, ,
\end{equation}
where we have denoted the interaction parameter as $\beta=q \,Q/m$. $Q$ and $q$ are the charge of the black hole and the charge of the particle, respectively. $m$ is the mass of the particle. 
%%%%%%%%%%%%%%
\begin{figure*}[!htb]
    \centering
    \includegraphics[scale=0.45]{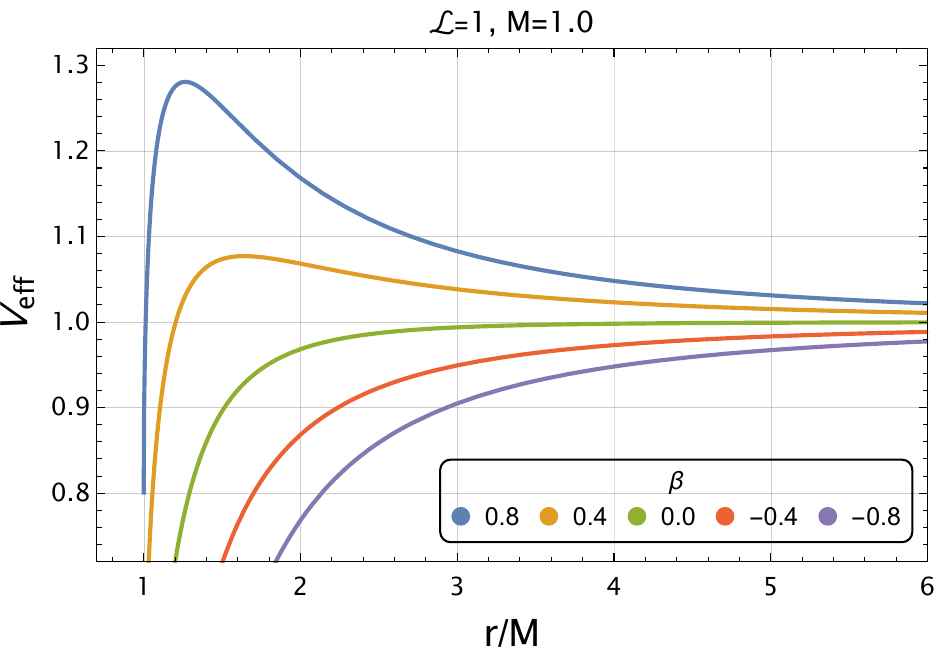}
    \includegraphics[scale=0.45]{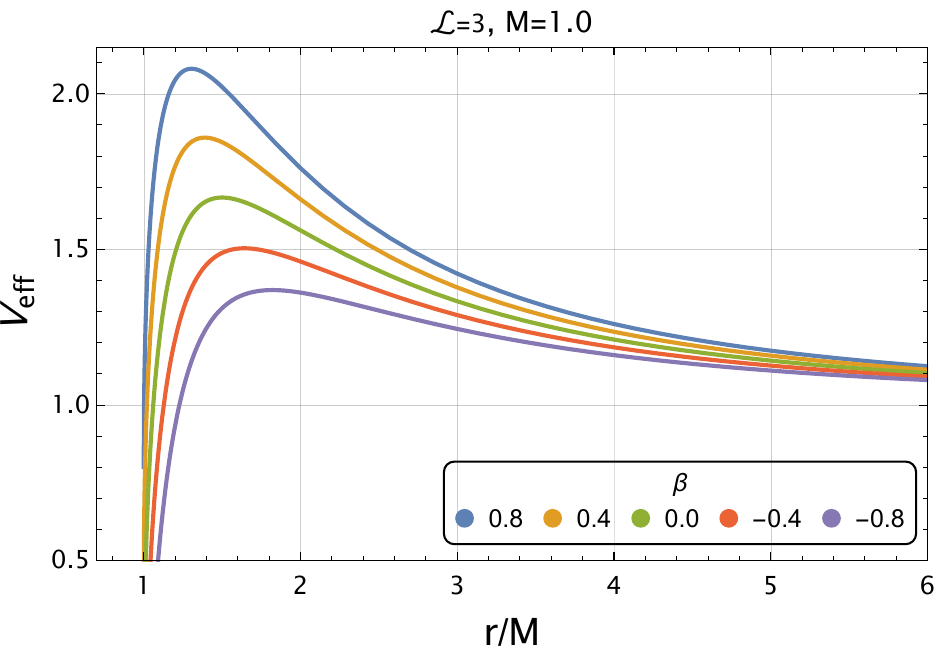}
    \caption{ The effective potential of a particle plotted against the distance $r$ from the black hole for $\mathcal{L}=1$ (left) and $\mathcal{L}=3 $(right).  }
    \label{fig:veff}
\end{figure*}
Fig. \ref{fig:veff} illustrates the effective potential for a charged particle around a five-dimensional weakly charged Schwarzschild black hole as a function of the radial distance  $r$. The figure shows that both $\beta$ and $\mathcal{L}$ have a similar influence on the effective potential. As $\beta$ and $\mathcal{L}$ increase, the height of the effective potential rises, and the peak shifts slightly to the left (towards smaller $r$).

\section{ELECTRIC PENROSE PROCESS IN 5D METRIC}\label{Sec:EP}

Let us assume that the particle is moving in the equatorial plane. By making this assumption, we aim to simplify the equations, resulting in the following simplified form for the five velocity components of the particle
\begin{equation} \label{five:velocity}
u^{\alpha} = u^t (1, v, 0, 0, \omega),
\end{equation}
where $v = dr/dt$, $\omega = d\chi/dt$, $d\theta/dt = d\phi/dt = 0$. 

Let us now write the normalization condition for the five velocity components of this particle
\begin{equation}
    u^\alpha u_\alpha = u^\alpha g_{\alpha\beta}u^\beta = -k .
\end{equation}
Using Eqs. \eqref{eq:5dmetric} and \eqref{five:velocity}, the normalization condition can be written as follows
\begin{equation} \label{norm:velocity}
(u^t)^2 (- f(r)+f(r)^{-1} v^2 + \omega^2 r^2) = -k ,
\end{equation}
where $k = 1$ for massive particle and $k = 0$ for photon.
If we solve Eq. \eqref{norm:velocity} with respect to $\omega$, we obtain the following expression for  the angular velocity  $\omega$ measured by a static observer at infinity
\begin{equation} \label{ang:velocity}
\omega = \pm \frac{1}{u_t r} \sqrt{(u_t)^2 [f(r) - f^{-1}(r) v^2] - kf^2(r)}.
\end{equation}
Given that the radial velocity $v =dr/dt = 0$ for a circular orbit and $k = 0$ for a photon, we conclude that the limiting values of angular velocity $\omega$ correspond to a photon moving in a circular orbit
\begin{equation}\label{ang:velrestrict}
\omega_- \le \omega \le \omega_+,
\end{equation}
where
\begin{equation} \label{ang:velrestrictvalue}
\omega_{\pm} = \pm \frac{\sqrt{f(r)}}{r}.
\end{equation}

\begin{figure*}[!htb]
    \centering
    \includegraphics[scale=0.45]{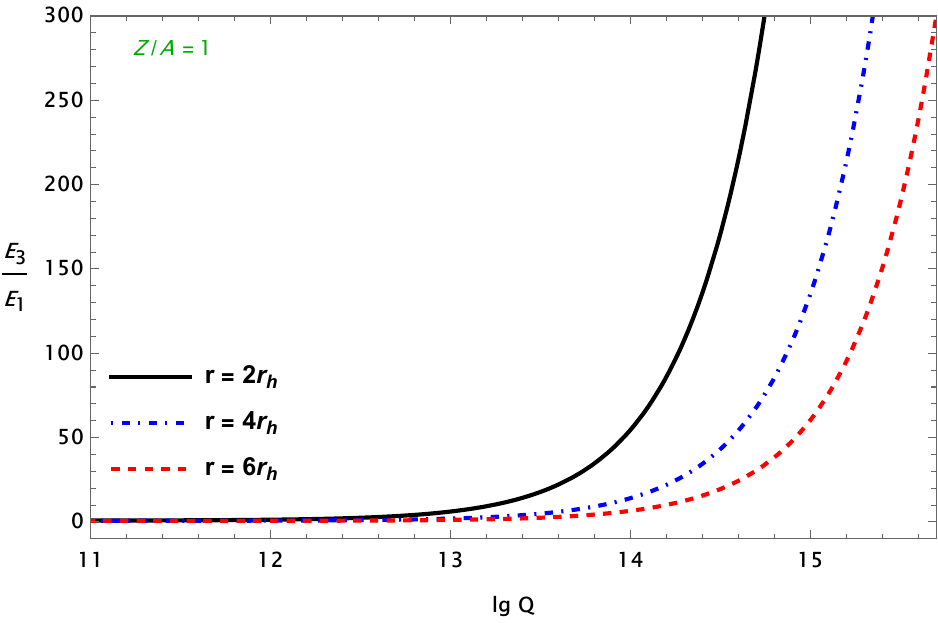}
    \includegraphics[scale=0.45]{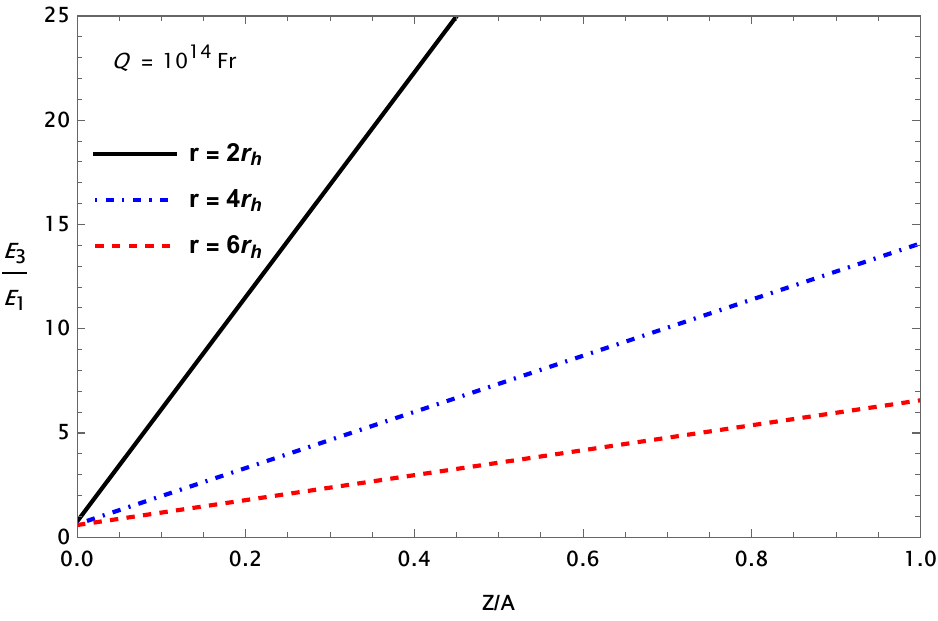}
    \includegraphics[scale=0.5]{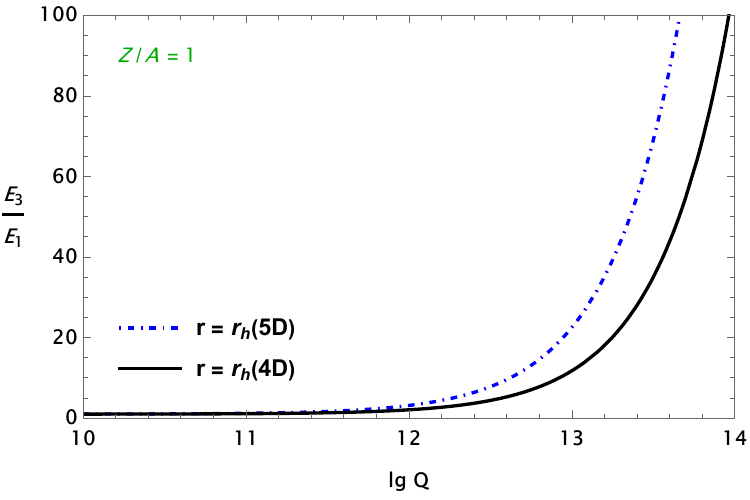}
    \caption{ (Top) Ratio of energies of ionized and neutral particles plotted against the black hole charge $Q$ (left) and the ionization rate $Z/A$ (right). The curves denote different positions of the ionization point, where $r_h = \sqrt{GM/c^2}$ is the horizon of the 5D Schwarzschild black hole. (Bottom) Comparison of ratio of energies of ionized and neutral particles around 4D (black, solid) and 5D (blue, dot-dashed) Scharzschild black holes plotted against the black hole charge $Q$ at each of their event horizons  $r_{h}(4D) = 2GM/c^2$ and $r_{h}(5D) = \sqrt{GM/c^2} $ respectively. }
    \label{fig:efficiency}
\end{figure*}

% \subsubsection{Conservation laws}

Now, we turn to deriving the expressions for the energy of the particle. Let's imagine, a neutral atom is falling into the weakly charged five-dimensional Schwarzschild black hole in the equatorial plane. It ionizes near the event horizon of the black hole. This means it splits into two fragments. In this case, we can write the following conservation laws
\begin{equation} \label{con:energy}
    E_1 = E_2 + E_3 ,
\end{equation}
\begin{equation} \label{con: ang-momentum}
    L_1 = L_2 + L_3 , 
\end{equation}
\begin{equation} \label{con: charge}
    q_1 = q_2 + q_3,
\end{equation}
\begin{equation} \label{con: momentum}
    m_1\dot{r_1} = m_2\dot{r_2} + m_3\dot{r_3} ,
\end{equation}
\begin{equation} \label{con: mass}
    m_1 \ge m_2 + m_3 ,
\end{equation}
where dot indicates derivatives with respect to the particle's proper time $\tau$. Note that physical quantity with lower index 1 corresponds to the falling particle. Indexes 2 and 3 correspond to two fragments after ionizing. Using the Eq. \eqref{con: momentum}, we can get the following equation \cite{Bhat85,Shaymatov24MPP1,Tursunov21EPP}
\begin{equation}\label{con:mom}
    m_{1}u_{1}^{\chi}=m_{2}u_{2}^{\chi}+m_{3}u_{3}^{\chi} \, .
\end{equation}

\begin{figure}[!htb]
    \centering
    \includegraphics[scale=0.4]{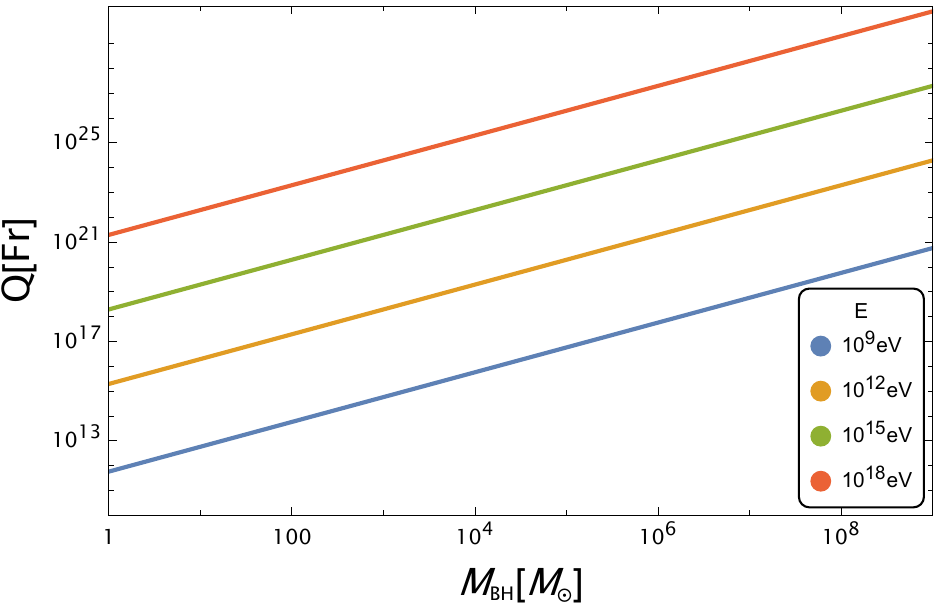}
    \includegraphics[scale=0.4]{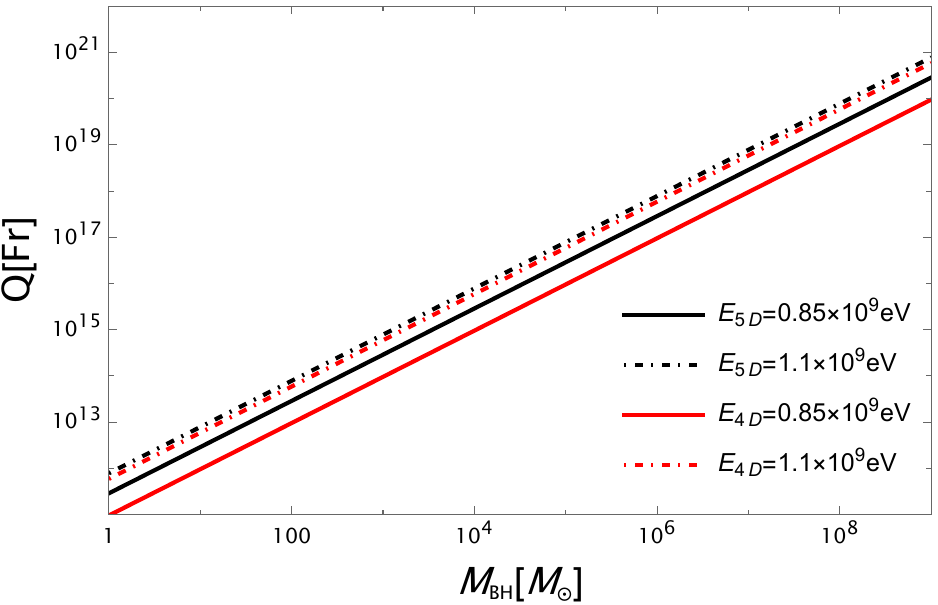}
        \caption{(Left) The relationship between the black hole mass and its charge for protons. Its energy varies from $1 \text{GeV}$ to $10^{12} \text{GeV}$). The ionization distance from the black hole is $r_i(5D) =  2\sqrt{GM/c^2}$.
        (Right) The comparison of the acceleration capability of 4D and 5D black holes for different energies of protons. The ionization distances from the black hole are $r_i(4D) = 4GM/c^2$ and $r_i(5D)= 2\sqrt{GM/c^2}$.
        }
    \label{fig:QM}
\end{figure}
\begin{figure*}[!htb]
    \centering
    \includegraphics[scale=0.45]{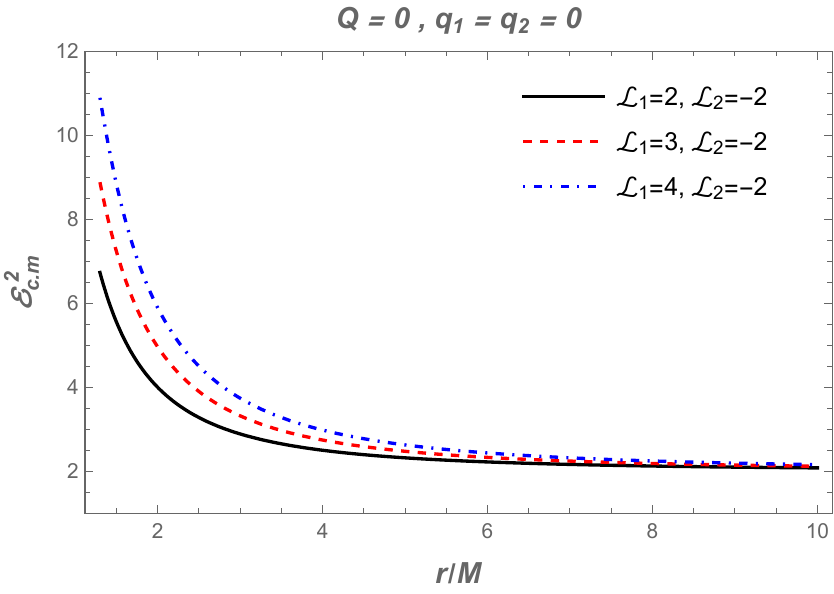}
    \includegraphics[scale=0.45]{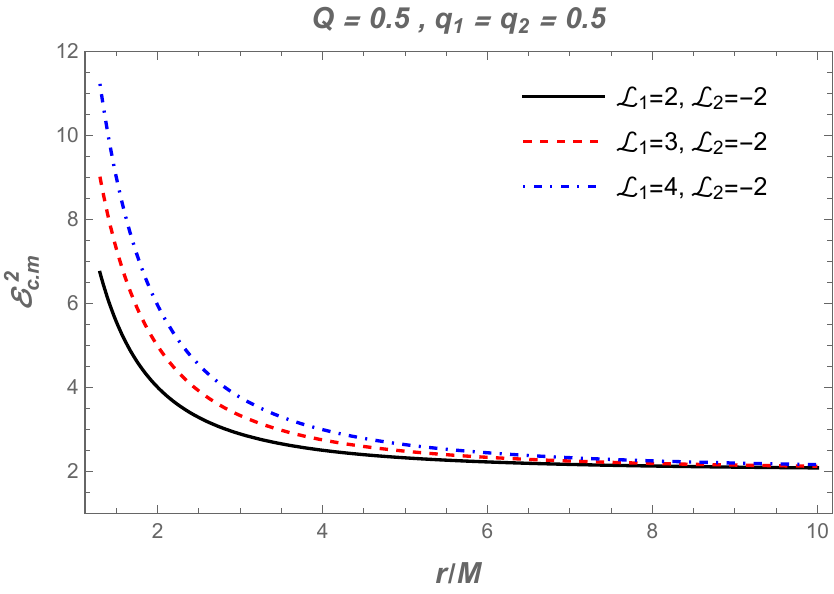}
    \includegraphics[scale=0.45]{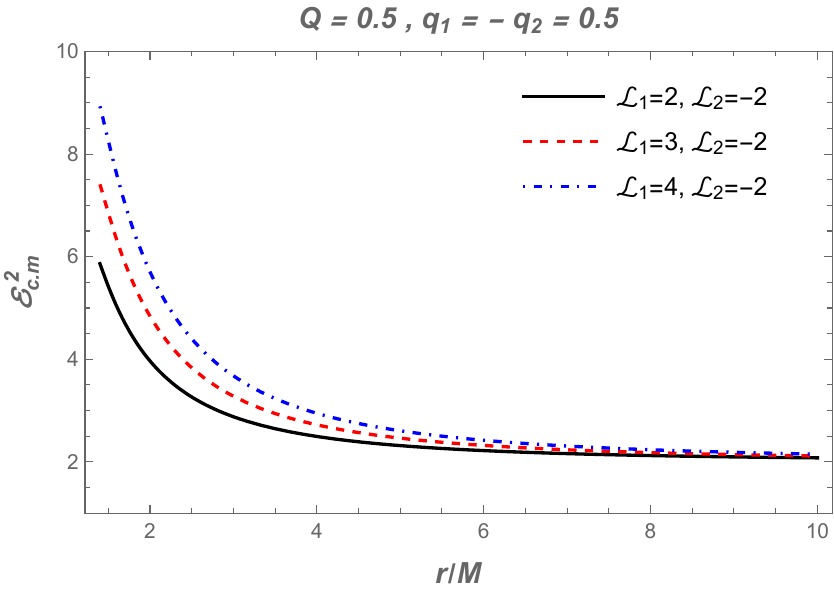}
  \includegraphics[scale=0.45]{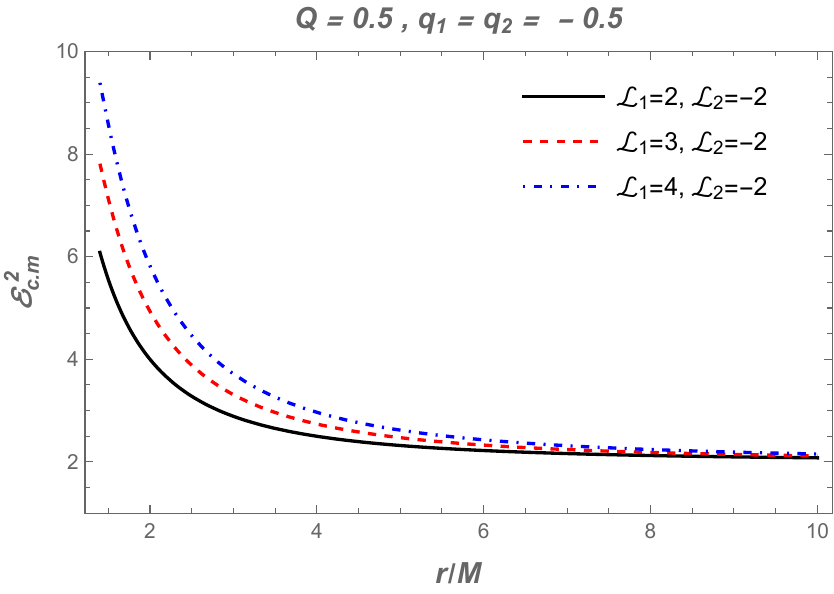}
    \caption{The relationship of ${\cal E}_{\rm c.m}^2$ to radial coordinate $r/M$ is described for different ${\cal{L}}_1$ and ${\cal{L}}_2$ values (upper panel).}
    \label{Fig:cm}
\end{figure*}
From Eqs. \eqref{five:velocity} and \eqref{int: motion-energy}, it is clear that  
\begin{equation} \label{five-vel-exp}
    u^{\chi}_{i}=\omega u^{t}_{i}= \frac{\omega e_{i}}{g_{tt}} = \frac{\omega e_{i}}{f(r)} , 
\end{equation}
where, $e_i=(E_i+q_i A_t)/m_i$, with $i=1,2,3$ indicating the particle's number. By substituting Eq. \eqref{five-vel-exp} into Eq. \eqref{con:mom}, we obtain the following expression:
\begin{equation}\label{con:momentum2}
    m_{1}\omega_{1}e_{1}=m_{2}\omega_{2}e_{2}+m_{3}\omega_{3}e_{3} .
\end{equation}
By solving the above equation for $E_3$, we can determine the energy of one of the particles formed after the ionization
\begin{equation} \label{energy:e3}
    E_{3}=\frac{\omega_{1}-\omega_{2}}{\omega_{3}-\omega_{2}}(E_{1}+q_{1}A_{t})-q_{3}A_{t} ,
\end{equation}
where $\omega_i=d\chi_i/dt$ is $i$ th particle's angular velocity. It is clear from equation (\ref{ang:velrestrict}), $\omega$ takes values between $\omega_{-}$ and $\omega_{+}$.

 Let us assume that the initial particle (e.g., particle 1) is neutral and at rest. Then the following equations are satisfied for this particle
\begin{align} \label{con:laws}
    q_{1} =0,\quad  E_{1} =m_1, \quad  u^t_1=1, \quad v = 0 .
\end{align}
Based on the above conditions and using Eq. \eqref{ang:velocity}, we define the angular velocity $\omega_1$ of particle 1, which takes the following form:
\begin{equation} \label{omega1}
   \omega_{1}=\frac{\sqrt{f(r)(1-f(r))}}{r} .
\end{equation}
For the energy $E_3$ (see Eq. \eqref{energy:e3}) of particle 3 formed from the ionization to reach its maximum value, the ratio $(\omega_{1}-\omega_{2})/(\omega_{3}-\omega_{2})$ must also reach its maximum. For this, the following conditions must be met:
\begin{equation}
    \omega_2 = \omega_{-} \quad \text{and} \quad 
    \omega_3 = \omega_{+} \ .
\end{equation}
In this case, the ratio becomes
\begin{equation} \label{max:omegas}
   \frac{\omega_{1}-\omega_{2}}{\omega_{3}-\omega_{2}}\Bigg|_{max}= \frac{\omega_{1}-\omega_{-}}{\omega_{+}-\omega_{-}} =\frac{1}{2}\sqrt{\frac{M} {r_{ion}^2}}+\frac{1}{2} ,
\end{equation}
where $r_{ion}$ is the ionization radius.

It is clear that expression (\ref{max:omegas}) is inversely proportional to the ionization radius. Therefore, the closer the ionization occurs to the event horizon, the larger the value of the expression becomes.
Let us rewrite the energy of the ionized particle
\begin{equation} \label{energy: e3max}
   E_{3}= \left(\frac{1}{2}\sqrt{\frac{GM} {c^2r_{ion}^2}}+\frac{1}{2} \right) E_{1}-q_{3}A_t .
\end{equation}
The energy of an ionized particle depends on the sign of the second term. If this term is negative, $E_3$ increases. For this to happen, the signs of $q_3$ and  the charge of the black hole $Q$ must be the same. In this case, the ionized particle is accelerated due to the Coulombic repulsion force between the black hole and the ionized particle.

Next, we determine the efficiency of the process by calculating the ratio of the accelerated particle's energy to the initial particle's energy. Let us imagine some neutral atom with the atomic number $Z$ and the mass number $A$ is completely ionized near a black hole. In this case, the charge of the ionized particle is $q_3 = Ze$ and the energy of initial particle is $E_1 = m_1 c^2  \approx A~m_{n}c^2$. Here, $e$ is the elementary charge and $m_n$ is nucleon mass. So, the efficiency takes the following form
\begin{equation} \label{energy:eff}
    \frac{E_{3}}{E_{1}}=\frac{1}{2}+\frac{1}{2}\sqrt{\frac{GM} {c^2r_{ion}^2}}+\frac{Ze Q}{A~m_{n}c^{2}r_{ion}^2} .
\end{equation}
This process is efficient only if the ratio $E_3/E_1$ exceeds unity. 

Thus, the region where energy can be extracted from a black hole with a positive charge $Q$ is given by $r_h<r_{ion}<r_{max}$. Here, $r_{max}$ is determined by
\begin{equation}
    r_{max} = \frac{1}{2}\left(\sqrt{\frac{GM} {c^2}}+\sqrt{\frac{GM} {c^2}+\frac{8Ze Q}{A~m_{n}c^{2}}}\right).
\end{equation}

When ionization occurs very close to the event horizon of the black hole ($r_h = \sqrt{G M/c^2}$), the energy of the ionized particle, $E_3$, will be greater than that of the initial particle, $E_1$, for any positive black hole charge, $Q > 0$. When ionization occurs at $2r_h$, the black hole's charge must satisfy the following condition to ensure that the ionized particle's energy ($E_3$) is greater than the energy of the neutral particle ($E_1$).
\begin{equation} \label{min:charge}
    Q\ge1.16\times10^{11}\frac{A}{Z}\frac{M}{M_{\odot}}\ \text{Fr} .
\end{equation}
This value is slightly above the lower limit of the estimated realistic limits for the black hole's charge \cite{zajek2019, Zajek_2018}.
In Fig. \ref{fig:efficiency} (top, left) we can see the energy efficiency of the electric Penrose process (ratio of energies of ionized ($E_3$) and neutral ($E_1$) against the charge of the black hole for different ionization points. The graph shows that with increasing the charge of the black hole, the efficiency grows fast. If the ionization occurs at a greater distance from the black hole, the efficiency decreases noticeably. In Fig. \ref{fig:efficiency} (top, right) we plot the ratio of energies of ionized and neutral particles against the ionization rate $Z/A$ for a fixed charge of the black hole ($Q = 10^{14}Fr$). In the graph the efficiency decreases considerably if the ionization occurs far away from the black hole. In Fig. \ref{fig:efficiency} (bottom) we compare acceleration capability of 4D and 5D Schwarzschild black holes. The graph is plotted the energy efficiency of two types of black holes against their charges. In the graph the ionization occurs near the horizon of each black hole. We can see acceleration capability of the 5D Schwarzschild black hole slightly better than 4D one.

In Fig. \ref{fig:QM} (top) we plot the charge of the black hole against the mass of the black hole for different values of  energies of the proton. The graph shows that more charge is required for increasing the energy of the proton. In Fig. \ref{fig:QM} (bottom) we compare the energies of the proton which is accelerated by 4D and 5D Schwarzschild black holes. One can see that less charge is required to get low-energy protons for 4D black hole than 5D one. However, the difference of the required charge to accelerate protons for 4D and 5D black holes decreases if the acceleration energy slightly increases.

\section{Collisions of electrically charged particles
near the event horizon of the 5D Schwarzschild black holes} \label{Sec:energy}

In this section, we study the collisions of the electrically charged particles near the event horizon of the 5D Schwarzschild BH.
The acceleration of particles colliding near rotating Kerr black holes has been first studied in ~\cite{Banados09}, where the center-of-mass energy of colliding particles may diverge in the case of the extreme rotating Kerr black hole.
Up to now, several authors have investigated the impact of external magnetic fields on the acceleration processes of charged particles in the vicinity of black holes in various gravity models and scenarios (see, for example, Refs.~\cite{Frolov12,Abdujabbarov13a}). It has been demonstrated that the efficiency of the energy extraction mechanism is more effective in head-on collisions.
The expression for the center of mass energy for two particles can be found as a sum of two-momenta ~\cite{Grib11a}
\begin{eqnarray}\label{cmen1}
\{E_{cm},0,0,0\}=m_1u_1^{\mu}+m_2u_2^{\mu}\ ,
\end{eqnarray}
where $u_1^{\alpha}$ and $u_2^{\beta}$ are the four-velocity of the two colliding particles with the masses $m_1$ and $m_2$, respectively. One can easily calculate the square of center of mass energy defined in (\ref{cmen1}) and get
\begin{eqnarray}
E^2_{cm}=m_1^2+m_2^2-2m_1m_2g_{\mu\nu} u^{\mu}u^{\nu}\ ,
\end{eqnarray}
or
\begin{equation}
\frac{E^2_{cm}}{m_1m_2}=\frac{m_1}{m_2}+\frac{m_2}{m_1}-2g_{\mu\nu} u^{\mu}_1u^{\nu}_2\ .
\end{equation}
In case when the masses of the colliding particles are $m_1=X m$ and $m_2=Y m$,  
\begin{equation}\label{ecm2}
\frac{E^2_{cm}}{m^2c^4}=X^2+Y^2-2g_{\mu\nu} u^{\mu}_1u^{\nu}_2\ .
\end{equation}
Following that, we will analyze the collision of particles of the same masses $m_1=m_2=m$ and initial energies $E_1=E_2=m$. Using the standard equation for the center-of-mass energy of two colliding particles with the same mass, we will investigate the acceleration of charged particles near the 5D Schwarzschild BH. As a result, the expression for the center-of-mass energy becomes
\begin{eqnarray}\label{ECMeq}
{\cal E}_{\rm c.m}^2=\frac{E^2_{\rm cm}}{2m^2 c^4}=1-g_{\alpha \beta}u_1^{\alpha}u_2^{\beta}\ ,
\end{eqnarray}
As a result, using the components of the four-velocity, the final expression for the center of mass energy in the equatorial plane (where $\theta=\pi/2$) has the following form:
%\begin{widetext}
\begin{eqnarray}\label{Eq:cmen}
\nonumber
&&{\cal E}^2_{\rm c.m}=1+\frac{1}{f(r)}\left({\cal E}_1-q_1 A_t \right)\left({\cal E}_2-q_2 A_t \right)-\\ 
&&-\frac{\mathcal{L}_1{\cal L}_2}{r^2} -\frac{1}{f(r)}\sqrt{\left({\cal E}_1-q_1 A_t\right)^2-f(r)\Big(1+\frac{{\cal L}_1^2}{r^2}\Big)}\sqrt{\left({\cal E}_2-q_2 A_t\right)^2-f(r)\Big(1+\frac{{\cal L}_2^2}{r^2}\Big)} \, .
\end{eqnarray}
Using Eq.~(\ref{Eq:cmen}), we can plot the radial dependence of the center of mass energy ${\cal E}^2_{\rm c.m}$ for different values of the parameters in Fig.~\ref{Fig:cm}. We can see from this figure that the value of the center of the mass energy decreases with the increase of the distance. It is clear from the figure that the value of the center of the mass energy increases under the influence of the ${\cal L}_1$ for a fixed value of the ${\cal L}_2$. Also, the figure is plotted for various combinations of the charge of particles. Moreover, there is a slight increase with the increase of the charge $Q$ in the right-bottom panel of Fig.~\ref{Fig:cm}.

\section{Conclusions}
\label{Sec:Conclusion}

In this paper, we studied a simple mechanism of energy harvesting from the weakly charged five-dimensional Schawarzschild black hole (its charge does not effect on the spacetime metric) and the collisions of charged particles near its event horizon. 
First, we considered the effective potential of the 5D black hole and its properties. 
We showed that the effective potential increases (decreases) as its electric interaction parameter $\beta$ (or angular momentum $\mathcal{L}$) increases (decreases). 
We have discovered that if the charges of the ionized particle and the black hole have the same sign, the ionized particle's energy can be far higher than the neutral particle's initial energy.
Also, we showed that the five-dimensional Schwarzschild black hole has better acceleration capability than its four dimensional counterpart if ionization occurs near their event horizons.
It is noted that slightly more charge is required for 5D black hole than its 4D counterpart to accelerate protons for the same energy if ionization happens far from the horizons of them. 
It is because of $r^2$, which is the denominator of the gravitational and electromagnetic four potential terms. This causes the potentials to decrease rapidly with increasing distance around 5D black hole.
Moreover, we have explored the collisions of electrically charged particles near the event horizon of the 5D Schwarzschild BH. 
We have calculated the center of mass energy using the appropriate equation and plotted the radial dependence for different values of the ${\cal L}$ and the charge of particles. The results are demonstrated in Fig.~\ref{Fig:cm}. 
We can see from this figure that the values of the center of mass energy increase with the increase of the first particle's angular momentum ${\cal{L}}_1$ while the angular momentum of the second particle ${\cal{L}}_2$ remains unchanged. 
Furthermore, one can see the behavior of the center of mass energy for different values of the charge of the particles in Fig.~\ref{Fig:cm}.

\section{Acknowledgments}
The authors would like to thank the anonymous referee for useful comments and suggestions, which
improved the quality of the manuscript.

\vspace{1.0 cm}
\section*{Data Availability Statement}
%\vspace{0.5 cm}
No Data associated in the manuscript. 

\bibliographystyle{unsrt}
\bibliography{gravreferences,ref1,ref2}

\end{document}